\RequirePackage{lineno}
\documentclass[%
 reprint,
showpacs,preprintnumbers,showkeys,
 amsmath,amssymb,
 aps,
prd,
floahttps://www.overleaf.com/project/62b96170c59dc8710ffcf234tfix,
]{revtex4}

\usepackage{graphicx}
\usepackage{dcolumn}
\usepackage{bm}
\usepackage{amssymb}
\usepackage{float}
\usepackage[caption = false]{subfig}
\usepackage{hyperref}

\begin{document}
\title{Intrinsic coupling between transverse spherocity and elliptic flow in heavy-ion collisions}

\author{Subikash Choudhury}
\email{subikash.choudhury@cern.ch}
\affiliation{Nuclear and Particle Physics Research Centre, Department of Physics, Jadavpur University, Kolkata - 700032, India}

\author{Ekata Nandy}
\email{ekatanandy@gmail.com}
\affiliation{Variable Energy Cyclotron Centre, 1/AF, Bidhan nagar, Kolkata - 700064, India}

\date{\today}

\begin{abstract}
Transverse spherocity ($S_{0}$) is an event-shape observable widely used to classify collision events according to their topology, particularly to distinguish jet-like from isotropic events. Low-spherocity events are generally interpreted as being associated with enhanced jet activity. This event classification has also been applied to heavy-ion collisions to investigate the influence of event topology on several observables, including elliptic flow and constituent-quark-number scaling~\cite{sph_HI_JPG48_2021,sph_HI_arxiv_2207_12133,sph_HI_epjc82_2022,sph_HI_epja_58_2022,sph_HI_SciRep}. In this work, we demonstrate that such an interpretation requires careful reconsideration. Using toy Monte Carlo simulations, A Multiphase Transport (AMPT) model calculations, and an analytical formulation of the spherocity observable, we show that transverse spherocity is intrinsically related to the elliptic flow coefficient, $v_{2}$. This connection arises because the axis that minimizes the spherocity aligns with the event symmetry plane, causing events with larger elliptic anisotropy to naturally exhibit smaller spherocity values even in the absence of genuine jet-like topologies. We further show that this intrinsic relation gives rise to an inherent anti-correlation between transverse spherocity and $v_{2}$, implying that several characteristics previously attributed to the enhanced jet-like nature of low-spherocity events in heavy-ion collisions can instead be understood as consequences of collective anisotropic flow. Our results indicate that, in heavy-ion collisions, transverse spherocity should be interpreted primarily as a probe of the collective momentum-space anisotropy rather than as a direct measure of jetty event topology. Consequently, physics conclusions drawn from spherocity-selected events should explicitly account for its intrinsic correlation with elliptic flow.
\end{abstract}


\maketitle


\section{Introduction}
The event-by-event characterization of relativistic heavy-ion collisions has become an important approach for studying the properties of the strongly interacting quark--gluon plasma (QGP). Fluctuations in the positions of the participating nucleons lead to event-by-event variations in the initial collision geometry, which are subsequently converted into anisotropies in the final-state momentum distributions through the collective expansion of the produced medium~\cite{intro_prl_103_2009,intro_prc_81_2010,intro_prc_91_2015,intro_alice_prl_107_2011,intro_phenix_prl_107_2011,intro_atlas_prc_86_2012,intro_atlas_epjc_78_2018}. Consequently, collisions with similar global properties can exhibit significantly different final-state momentum distributions. This has motivated the development of event-classification techniques that group collisions according to their global characteristics, allowing a more detailed investigation of the underlying collision dynamics.

The most commonly used event classifier is the collision centrality, which categorizes events according to the degree of geometrical overlap between the colliding nuclei. While centrality provides a convenient measure of the overall collision geometry, events belonging to the same centrality class can still possess substantially different initial-state eccentricities and collective responses. To exploit these residual event-by-event fluctuations, several event-shape observables have been introduced. Among them, the reduced flow vector, (q$_n$), forms the basis of Event Shape Engineering (ESE), where events are classified according to the magnitude of their flow anisotropy, enabling systematic investigations of the relationship between the initial geometry and the collective evolution of the medium~\cite{jshukraft_plb_2012,atlas_ese_prc_2015,alice_ese_prc_2016}. The success of this approach naturally motivates the exploration of other event-shape observables that may offer complementary information about the collision dynamics.

One such observable is the transverse spherocity ($S_0$), originally introduced in high-energy elementary collisions as an infrared- and collinear-safe event-shape variable that quantifies the geometrical distribution of energy-momentum flow~\cite{abanfi_jhep_2010}. In proton--proton and $e^{+}e^{-}$ collisions, transverse spherocity effectively distinguishes pencil-like dijet events from isotropic multihadronic events. Events with small values of $S_0$ correspond to highly collimated back-to-back jet topologies, whereas values approaching unity indicate nearly isotropic momentum distributions. Owing to these properties, event-shape observables such as spherocity and sphericity have played an important role in studies of perturbative Quantum Chromodynamics, including precision determinations of the strong coupling constant $\alpha_s$~\cite{alphaS_jade_epjc_1998,mdasgupta_jpg_2003,alphaS_hadr_opal_epjc_2005,alphaS_hadr_jade_epjc_2009,alphaS_hadr_opal_epjc_2011}.

More recently, transverse spherocity based measurements has been carried out in pp collisions at LHC energies ~\cite{sph_pp_alice_epjc_2012,aotiz_arxiv1404,aotiz_npa_2015,sph_pp_alice_epjc_2019} and extended to heavy-ion collisions with the objective of separating jet-like and isotropic events in a manner analogous to pp collisions~\cite{sph_HI_JPG48_2021,sph_HI_arxiv_2207_12133,sph_HI_epjc82_2022,sph_HI_epja_58_2022,sph_HI_SciRep}. Some of these studies have reported that events selected with low-S$_{0}$ exhibit enhanced elliptic flow, stronger radial expansion, and a larger violation of constituent-quark-number (NCQ) scaling of the elliptic flow coefficient. These observations have generally been interpreted as evidence that low-$S_0$ events are preferentially enriched with jet-like topologies, while high-$S_0$ events represent predominantly soft and isotropic particle production.

However, whether this interpretation remains valid in the environment of heavy-ion collisions requires careful examination. Unlike elementary collisions, where the event topology is primarily determined by hard partonic scatterings, the final state of a heavy-ion collision is dominated by collective expansion of a strongly interacting medium. As a consequence, the momentum distribution of produced particles possesses a well-defined symmetry axis associated with the reaction plane or event plane. Since the definition of transverse spherocity itself requires the determination of a direction that minimizes the transverse momentum projected perpendicular to a unit vector, an important question naturally arises- does the minimizing direction identify jet topology, or does it instead align with the symmetry plane of the event?

This distinction is crucial because, if the minimizing direction is controlled primarily by collective flow, transverse spherocity cannot be regarded as an independent measure of event topology in heavy-ion collisions. Instead, its value would be intrinsically correlated with the anisotropic flow coefficients, particularly the elliptic flow coefficient $v_2$. In such a scenario, the observed properties of low-$S_0$ events could arise naturally from their larger collective anisotropy rather than from enhanced jet production.

In this work, we investigate this question using three complementary approaches. First, we employ simple toy Monte Carlo simulations to illustrate the geometrical origin of the minimizing direction in the definition of transverse spherocity. We then validate these observations using events generated with the AMPT transport model. Finally, we derive an analytical relation between transverse spherocity and the second-order anisotropic flow coefficient, demonstrating that the two observables are intrinsically coupled through their common symmetry axis and exhibit an anti-correlation that follows naturally from the mathematical construction of the spherocity observable and does not require the presence of jet-like event topology. These findings suggest that the interpretation of low-$S_0$ events in heavy-ion collisions should be reconsidered and that the use of transverse spherocity as an event classifier must account for its inherent correlation with collective anisotropic flow.

\section{Toy model studies}

Before investigating realistic heavy-ion collisions, it is instructive to examine the behaviour of transverse spherocity in simple event topologies where the preferred direction of particle emission is known a priori. To this end, we construct a toy Monte Carlo (TOY MC) model consisting of idealized jet-like and isotropic events. The purpose of this study is twofold, first, to verify that the spherocity algorithm correctly identifies the preferred emission direction in a dijet topology, and second, to understand how its behaviour changes when the event topology is governed by anisotropic flow rather than jets.
\begin{figure}[htbp]
\centering
\includegraphics[width=120mm,height=5.5cm]{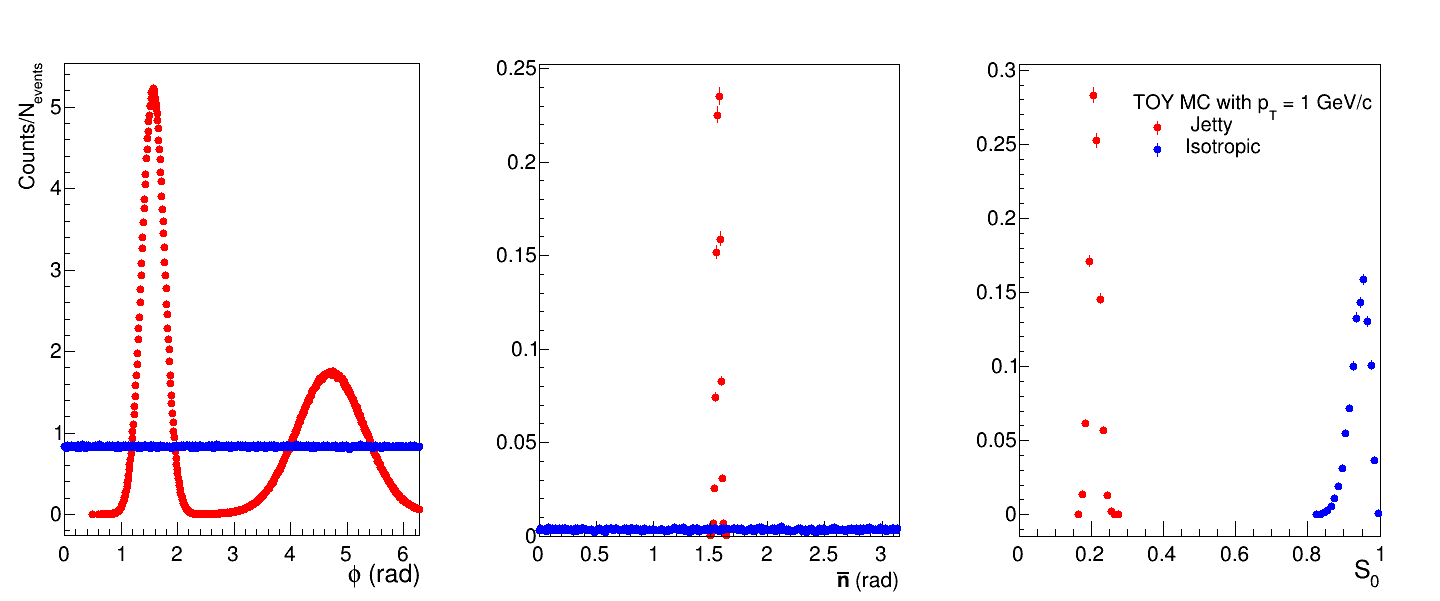}
	\caption{\label{TOY-MC-jetty-isotropic}[Color online] (Left) Azimuthal ($\phi$) distribution for dijet-like and uniform event topology shown in red and blue marker, respectively. (Middle) Phase angle distribution of unit vector, $\vec{n}$ that minimizes the quantity in Eq.~\ref{eqn1} for dijet-like (red) and uniform event topology (blue). (Right) EbyE distribution of transverse spherocity S$_{0}$ for dijet-like (red) and uniform (blue) event-topology. }
\end{figure}

\subsection{Jet-like and isotropic event topologies}

A jet-like event is simulated by generating particles according to two Gaussian distributions in the azimuthal angle ($\phi$), representing the near-side and away-side jets. The Gaussian distributions are centered at $\pi/2$ and $3\pi/2$, respectively, with the near-side jet chosen to have a narrower width than the away-side jet. The resulting azimuthal distribution is shown in the left panel of Fig.~\ref{TOY-MC-jetty-isotropic}. 
\begin{equation}
S_{0} =\frac{\pi^{2}}{4} min\left( \frac{\sum_{i}|\vec{p_{T,i}} \times \vec{n}|}{\sum_{i}\vec{|p_{T,i}|}} \right)^{2}
\label{eqn1}
\end{equation}
The middle panel of Fig.~\ref{TOY-MC-jetty-isotropic} shows the distribution of the minimizing direction, $\vec{n}$, that minimizes the transverse spherocity defined in Eq.~\ref{eqn1} . As expected, the distribution exhibits a sharp peak at $\phi=\pi/2$, demonstrating that the spherocity algorithm correctly identifies the jet axis as the preferred direction of particle emission. This confirms the conventional interpretation of transverse spherocity as an event-shape observable capable of identifying jet-like topologies.

For comparison, isotropic events are generated by distributing particles uniformly over the full azimuthal range. In the absence of any preferred emission direction, the minimizing axis is found to be uniformly distributed between 0 and $\pi$, as shown in the middle panel of Fig.~\ref{TOY-MC-jetty-isotropic}. The corresponding event-by-event distributions of $S_0$, shown in the right panel of the same figure, clearly separate the two event classes. Jet-like events populate the low-$S_0$ region, whereas isotropic events produce values close to unity, illustrating the ability of transverse spherocity to distinguish between collimated and isotropic particle production in systems resembling p-p collisions.
\begin{figure}[htbp]
\centering
\includegraphics[width=120mm,height=6.0cm]{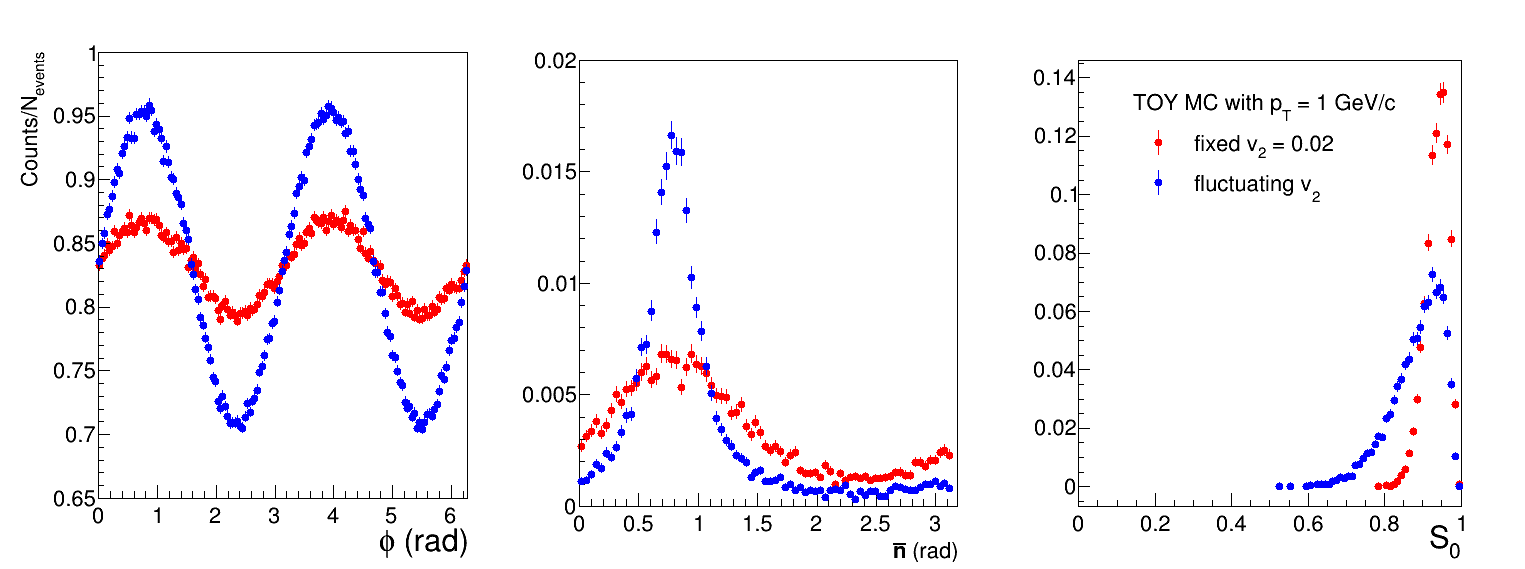}
	\caption{\label{TOY-MC-v2}[Color online] (Left panel) Flow modulated $\phi$-distribution
	for fixed value of v$_{2}$ (red marker) and fluctuating v$_{2}$ (blue marker). (Middle panel) Phase angle distribution of unit vector for fixed value of v$_{2}$ (red marker) and fluctuating v$_{2}$ (blue marker). (Right panel) Measure of S$_{0}$ for fixed value of v$_{2}$ (red marker) and fluctuating v$_{2}$ (blue marker).}
\end{figure}
\subsection{Flow-modulated event topology}

The situation is fundamentally different in relativistic heavy-ion collisions. Owing to the large particle multiplicity and the strong collective expansion of the medium, the global event topology is dominated by azimuthal anisotropic flow rather than by isolated jets. Furthermore, the large underlying event makes the identification of true jets increasingly difficult, particularly in the transverse-momentum region $p_{\mathrm{T}}\lesssim$10 GeV/c, where transverse spherocity is typically calculated. This raises an important question of what actually determines the preferred direction identified by the spherocity minimization?

We hypothesize that, in heavy-ion collisions, the minimizing direction is governed primarily by the collective azimuthal anisotropy rather than by jet activity. If this picture is correct, the unit vector $\vec{n}$ should align with the reaction plane (or equivalently the event plane in the ideal limit, along which particle emission is enhanced due to elliptic flow.

To test this conjecture, we generate events according to the azimuthal distribution

\begin{equation}
P(\phi)\propto1+2v_2\cos2(\phi-\Psi_{\mathrm{RP}}),
\end{equation}

where the elliptic flow coefficient is fixed at $v_2=0.02$ and the reaction-plane angle is chosen as $\Psi_{\mathrm{RP}}=\pi/4$. In a second set of simulations, event-by-event fluctuations are introduced by sampling $v_2$ from a Gaussian distribution centered at 0.02 while retaining only positive values.

The resulting azimuthal distributions are shown in the left panel of Fig.~\ref{TOY-MC-v2}. The corresponding distributions of the minimizing direction are presented in the middle panel. In both cases, the distribution exhibits a clear maximum at $\Psi_{\mathrm{RP}}=\pi/4$ demonstrating that the axis selected by the spherocity minimization coincides with the reaction plane. This behaviour is qualitatively different from the jet-like case, where the minimizing direction follows the jet axis. Instead, in a collective medium, the preferred direction is determined by the symmetry axis of the anisotropic particle distribution.

The corresponding $S_0$ distributions are shown in the right panel of Fig.~\ref{TOY-MC-v2}. Introducing event-by-event fluctuations in $v_2$ increases the probability of generating events with larger momentum-space anisotropies. Such events exhibit a stronger preference for particle emission along the reaction plane, resulting in systematically smaller values of transverse spherocity. Consequently, the $S_0$ distribution shifts towards lower values. This observation demonstrates that, in heavy-ion collisions, low values of transverse spherocity need not originate from jet-like topologies but can arise naturally from the collective anisotropic expansion of the medium.
\begin{figure}[htbp]
\centering
\includegraphics[width=120mm,height=7.0cm]{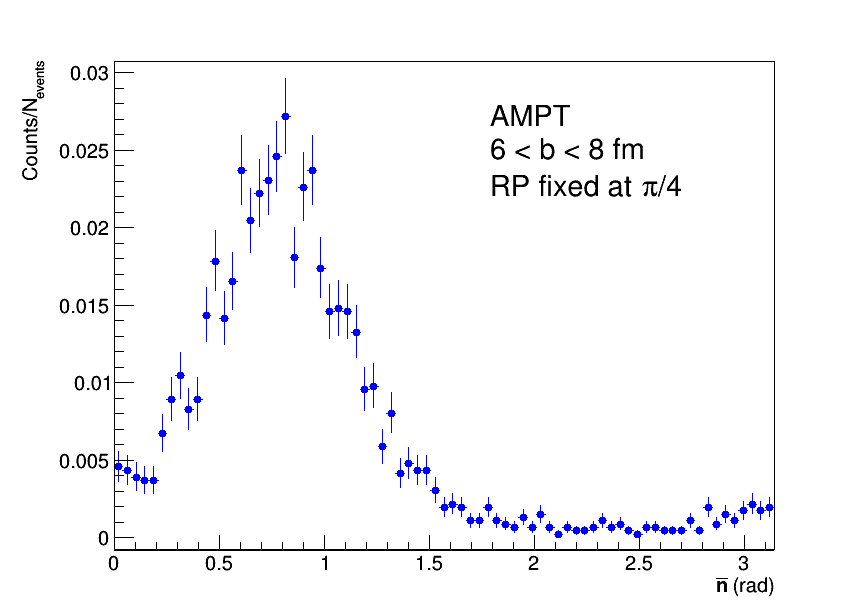}
\caption{\label{AMPT-n-unit} Phase angle distribution of unit vector obtained from events simulated with AMPT~\cite{ampt} model for Pb-Pb collisions in impact parameter range 6 $<$ b $<$ 8 fm at fixed reaction plane angle, $\Psi_{RP}$ = $\pi$/4  }
\end{figure}
To verify that this behaviour is not an artifact of the simplified toy model, we repeat the analysis using events generated with the AMPT model~\cite{ampt}. Several thousand Pb--Pb events are simulated in the impact parameter range $6<b<8$ fm while keeping the reaction-plane angle fixed at $\Psi_{\mathrm{RP}}=\pi/4$. The distribution of the minimizing direction obtained from the AMPT events is presented in Fig.~\ref{AMPT-n-unit}. Similar to the toy-model results, the distribution peaks at the imposed reaction-plane angle, confirming that the preferred direction identified by the spherocity algorithm is governed by the symmetry axis of the event. This agreement demonstrates that the toy model captures the essential physics underlying the behaviour of transverse spherocity in heavy-ion collisions and provides the basis for the analytical interpretation presented in the following section.

\section{Analytical formulation}

The numerical studies presented in the previous section suggest that the unit vector, $\vec{n}$, which minimizes the transverse spherocity in heavy-ion collisions aligns with the reaction plane. In this section, we provide an analytical framework to understand the origin of this behaviour and establish the connection between transverse spherocity and the azimuthal anisotropy of particle production.

The transverse spherocity defined in Eq.~\ref{eqn1} can be re-written as

\begin{equation}
S_{0}=
\frac{\pi^{2}}{4}
\left(
\min_{\psi}
\frac{\sum_{i}p_{T,i}\left|\sin(\phi_{i}-\psi)\right|}
{\sum_{i}p_{T,i}}
\right)^{2},
\label{eq:s0_definition}
\end{equation}

where $\psi$ denotes the phase angle of the unit vector $\vec{n}$ that minimizes the quantity inside the parentheses.

For high-multiplicity heavy-ion collisions, the discrete particle distribution may be approximated by a continuous azimuthal probability density $P(\phi)$ satisfying

\begin{equation}
\int_{0}^{2\pi}P(\phi)\,d\phi=1.
\end{equation}

To isolate the geometrical origin of the correlation between transverse spherocity and azimuthal anisotropy, we first consider equal particle weights i.e, $p_{T,i}=1$. Under this approximation, Eq.~(\ref{eq:s0_definition}) becomes

\begin{equation}
S_{0}
=
\frac{\pi^{2}}{4}
\left[
\min_{\psi}
F(\psi)
\right]^{2},
\label{eq:s0_continuum}
\end{equation}

where

\begin{equation}
F(\psi)
=
\int_{0}^{2\pi}
P(\phi)
\left|\sin(\phi-\psi)\right|
d\phi .
\label{eq:functional}
\end{equation}

The azimuthal distribution of particles produced in relativistic heavy-ion collisions can be expressed through the Fourier expansion

\begin{equation}
P(\phi)
=
\frac{1}{2\pi}
\left[
1
+
2\sum_{n=1}^{\infty}
v_{n}
\cos
n(\phi-\Psi_{n})
\right],
\label{eq:flow_distribution}
\end{equation}

where $v_{n}$ and $\Psi_{n}$ denote the magnitude and symmetry-plane angle of the $n^{\rm th}$ flow harmonic, respectively.

Similarly, the term $\left|\sin(\phi-\psi)\right|$ in Eq.~(\ref{eq:functional}) has the exact Fourier expansion as (details can be found in Appendix~\ref{app:fourier})

\begin{equation}
|\sin x|
=
\frac{2}{\pi}
-
\frac{4}{\pi}
\sum_{m=1}^{\infty}
\frac{\cos(2mx)}
{4m^{2}-1}.
\label{eq:kernel}
\end{equation}

Substituting Eqs.~(\ref{eq:flow_distribution}) and (\ref{eq:kernel}) into Eq.~(\ref{eq:functional}) yields

\begin{equation}
\begin{aligned}
F(\psi)
=
\frac{1}{2\pi}
\int_{0}^{2\pi}
&
\left[
1
+
2\sum_{n=1}^{\infty}
v_{n}
\cos n(\phi-\Psi_{n})
\right]
\\
\times
&
\left[
\frac{2}{\pi}
-
\frac{4}{\pi}
\sum_{m=1}^{\infty}
\frac{\cos2m(\phi-\psi)}
{4m^{2}-1}
\right]
d\phi .
\end{aligned}
\label{eq:expand}
\end{equation}

Using the orthogonality relation

\begin{equation}
\int_{0}^{2\pi}
\cos(n\phi)\cos(m\phi)
\,d\phi
=
\pi\delta_{nm},
\end{equation}

all mixed harmonic terms vanish, only even Fourier modes contribute to the integral. Consequently,

\begin{equation}
F(\psi)
=
\frac{2}{\pi}
-
\frac{4}{\pi}
\sum_{m=1}^{\infty}
\frac{
v_{2m}
\cos\left[2m(\Psi_{2m}-\psi)\right]
}
{4m^{2}-1}.
\label{eq:general}
\end{equation}

Equation~(\ref{eq:general}) establishes the analytical connection between transverse spherocity and the complete spectrum of even-order flow harmonics.

The preferred direction is obtained by minimizing Eq.~(\ref{eq:general}) with respect to $\psi$,

\begin{equation}
\frac{dF}{d\psi}
=
-
\frac{8}{\pi}
\sum_{m=1}^{\infty}
\frac{
m\,v_{2m}
\sin\left[2m(\Psi_{2m}-\psi)\right]
}
{4m^{2}-1}.
\label{eq:first_derivative}
\end{equation}

In relativistic heavy-ion collisions, the elliptic flow coefficient is typically much larger than the higher-order harmonics ($v_{2}\gg v_{4}>v_{6}\cdots$). Retaining only the dominant second harmonic, Eq.~(\ref{eq:first_derivative}) reduces to

\begin{equation}
\frac{dF}{d\psi}
\simeq
-
\frac{8}{3\pi}
v_{2}
\sin2(\Psi_{2}-\psi).
\end{equation}

The solutions corresponding to the extrema are given by, 

\begin{equation}
\psi=\Psi_{2},
\qquad
\psi=\Psi_{2}+\frac{\pi}{2}.
\end{equation}

The corresponding second derivative is

\begin{equation}
\frac{d^{2}F}{d\psi^{2}}
=
\frac{16}{3\pi}
v_{2}
\cos2(\Psi_{2}-\psi).
\end{equation}

For positive elliptic flow ($v_{2}>0$),

\begin{equation}
\left.
\frac{d^{2}F}{d\psi^{2}}
\right|_{\psi=\Psi_{2}}
>0,
\end{equation}

demonstrating that

\begin{equation}
\boxed{\psi_{\rm min}=\Psi_{2}.}
\end{equation}

Thus, in the continuum limit, the axis that minimizes the transverse spherocity naturally aligns with the second-order symmetry plane.

Substituting $\psi=\Psi_{2}$ into Eq.~(\ref{eq:general}) gives

\begin{equation}
F_{\rm min}
=
\frac{2}{\pi}
-
\frac{4}{\pi}
\sum_{m=1}^{\infty}
\frac{v_{2m}}
{4m^{2}-1}.
\end{equation}

Therefore,

\begin{equation}
S_{0}
=
\frac{\pi^{2}}{4}
\left[
\frac{2}{\pi}
-
\frac{4}{\pi}
\sum_{m=1}^{\infty}
\frac{v_{2m}}
{4m^{2}-1}
\right]^{2}.
\label{eq:master}
\end{equation}

When the elliptic component dominates the azimuthal anisotropy, the higher-order harmonics may be neglected, yielding

\begin{equation}
F_{\rm min}
\simeq
\frac{2}{\pi}
\left(
1-\frac{2}{3}v_{2}
\right),
\end{equation}

and consequently,

\begin{equation}
S_{0}
\simeq
\left(
1-\frac{2}{3}v_{2}
\right)^{2}.
\label{eq:approx}
\end{equation}

For the values of $v_{2}$ typically encountered in relativistic heavy-ion collisions ($v_{2}\lesssim0.1$), Eq.~(\ref{eq:approx}) may be expanded as

\begin{equation}
S_{0}
=
1
-
\frac{4}{3}v_{2}
+
\frac{4}{9}v_{2}^{2}
+\mathcal{O}(v_{2}^{3}),
\end{equation}

demonstrating that transverse spherocity decreases monotonically with increasing elliptic flow. It is worth noting that this behaviour is not unique to transverse spherocity. A closely related event-shape observable, transverse sphericity, also exhibits a monotonic anti-correlation with the elliptic flow coefficient under the factorization approximation. For completeness, the corresponding derivation is presented in Appendix~\ref{app:sphericity}.

\begin{figure}[htbp]
\centering
\includegraphics[width=120mm,height=6.0cm]{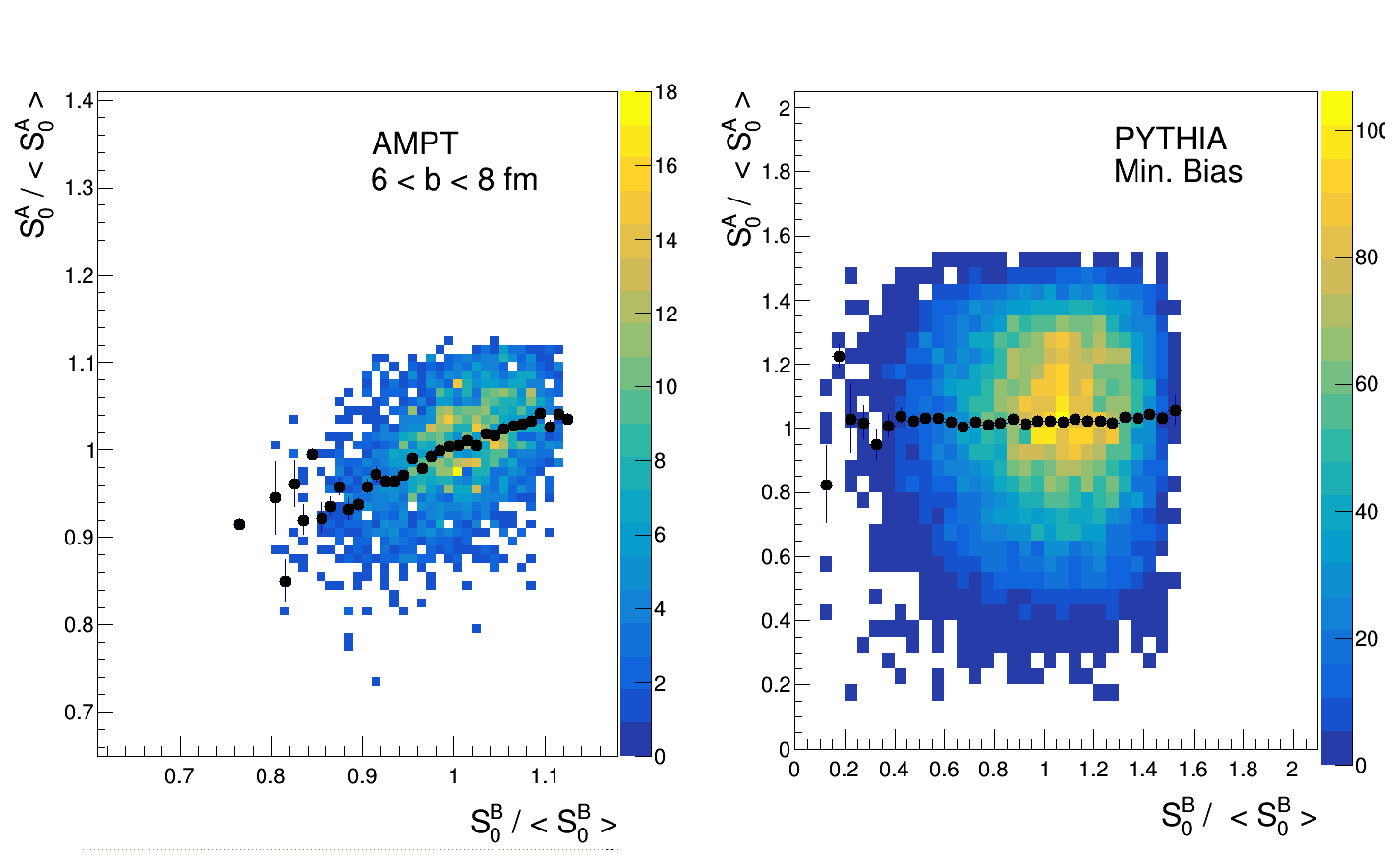}
\caption{\label{S0-correlation-AMPT-PYTHIA} Correlation between self normalized spherocity calculated in two disjoint $\eta$-range, A (-1 $< \eta <$ -0.5) and B (0.5 $< \eta <$ 1.0). (Left panel) Events simulated for Pb-Pb collisions in AMPT. (Right panel) PYTHIA simulation for pp collisions.}
\end{figure}
\subsection{Global nature of transverse spherocity in heavy ion collisions}
The analytical formulation presented in the previous section demonstrates that, in the continuum limit, the axis that minimizes the transverse spherocity aligns with the second-order symmetry plane of the event. Consequently, if transverse spherocity is indeed governed by the symmetry axis, it should exhibit long-range correlations across different regions of pseudorapidity($\eta)$. To test this expectation, Fig.~\ref{S0-correlation-AMPT-PYTHIA} shows the correlation between self-normalized $S_{0}$ values calculated in two mutually exclusive $\eta$-intervals. The quantities $S_{0}^{A}$ and $S_{0}^{B}$ are evaluated in the ranges $-1<\eta<-0.5$ and $0.5<\eta<1.0$, respectively. The left panel presents Pb--Pb collisions simulated with the AMPT model, while the right panel shows pp collisions generated with PYTHIA8~\cite{pythia8} with multiple parton interactions (MPI) enabled.

A strong positive correlation is observed between $S_{0}^{A}$ and $S_{0}^{B}$ for the AMPT events. Since particles produced in the two widely separated $\eta$ intervals originate from the same collision, they share a common event symmetry plane. As demonstrated by the analytical formulation, the minimizing direction of the spherocity is determined by this symmetry plane. Consequently, the transverse spherocity calculated in different $\eta$ intervals reflects the same underlying event geometry, causing $S_{0}$ to behave as a global event-shape observable in heavy-ion collisions.

In contrast, no significant correlation is observed for the PYTHIA events. In p--p collisions, where the event topology is dominated by localized particle production rather than collective flow, the minimizing direction is not constrained by a common symmetry plane. As a result, the transverse spherocity calculated in two disjoint $\eta$ intervals is statistically independent, leading to the absence of any appreciable correlation between $S_{0}^{A}$ and $S_{0}^{B}$. This comparison provides further evidence that the physical interpretation of transverse spherocity depends on the collision system: while it primarily characterizes local event topology in p--p collisions, in heavy-ion collisions it becomes a probe of the global collective geometry.

\section{Summary and Conclusion}

In this work, we have investigated the physical interpretation of transverse spherocity in relativistic heavy-ion collisions using analytical calculations together with toy-model and transport-model simulations. While transverse spherocity has traditionally been interpreted as an event-shape observable that discriminates between jet-like and isotropic topologies in p--p collisions, we demonstrate that this interpretation is no longer applicable in the presence of strong collective flow. Both the analytical formulation and numerical studies show that, in heavy-ion collisions, the axis that minimizes the transverse spherocity aligns with the second-order symmetry plane of the event. Consequently, transverse spherocity becomes intrinsically sensitive to the magnitude of the collective azimuthal anisotropy rather than to the presence of jets.

Within the continuum approximation, we derive an analytical relation connecting transverse spherocity to the even-order flow harmonics. When the elliptic flow dominates the azimuthal anisotropy, the leading-order expression predicts a monotonic decrease of $S_{0}$  with increasing $v_{2}$, providing a natural explanation for the strong anti-correlation observed between these two observables. Furthermore, the observed correlation of $S_{0}$  measured in two widely separated $\eta$-intervals demonstrates that, unlike in p--p collisions, transverse spherocity behaves as a global event-shape observable in heavy-ion collisions.

These findings provide a new perspective for interpreting several previously reported observations based on $S_{0}$-selected events. The pronounced anti-correlation between transverse spherocity and the reduced flow vector $q_{2}$ follows naturally from the analytical relation derived in this work, as both observables are governed by the same underlying elliptic flow. Consequently, many of the phenomena observed using $S_{0}$-based event selection are expected to resemble those obtained through Event Shape Engineering based on $q_{2}$. 

Our results also offer a different interpretation of the interplay between event shape and the collective expansion reported in Ref.~\cite{sph_HI_SciRep}. In that study, low-$S_{0}$ events were found to exhibit a larger average radial flow velocity $\beta$ together with a lower kinetic freeze-out temperature $T_{\mathrm{kin}}$. Similar behaviour has previously been observed using $q_{2}$-based Event Shape Engineering~\cite{alice_ese_prc_2016}, where the radial expansion was shown to be correlated with the initial spatial eccentricity. Since transverse spherocity is intrinsically correlated with the event-by-event elliptic flow, these observations can be understood as a consequence of the common underlying orientation of symmetry axis rather than as evidence of a jet bias introduced by the $S_{0}$-based event selection.

Overall, the present study demonstrates that the physical meaning of transverse spherocity depends strongly on the collision system. While it remains an effective discriminator between jet-like and isotropic topologies in small systems, p--p in particular, its interpretation in heavy-ion collisions is fundamentally different, where it predominantly reflects the underlying geometry and the associated anisotropic expansion of the medium. These findings establish a unified framework for understanding transverse spherocity across different collision systems and provide a basis for interpreting future measurements employing spherocity-based event classification in relativistic heavy-ion collisions.


\appendix

\section{Relation between transverse sphericity and elliptic flow}
\label{app:sphericity}

In this appendix, we derive the relation between transverse sphericity and elliptic flow under the factorization approximation. The linearized transverse sphericity tensor is defined as~\cite{sph_pp_alice_epjc_2012}

\begin{equation}
S^{L}_{XY}
=
\frac{1}{\sum_i p_{T,i}}
\sum_i
\frac{1}{p_{T,i}}
\begin{pmatrix}
p_{x,i}^{2} & p_{x,i}p_{y,i}\\
p_{y,i}p_{x,i} & p_{y,i}^{2}
\end{pmatrix}.
\label{eq:A1}
\end{equation}

Using

\[
p_{x}=p_{T}\cos\phi,
\qquad
p_{y}=p_{T}\sin\phi,
\]

Eq.~(\ref{eq:A1}) becomes

\begin{equation}
S^{L}_{XY}
=
\frac{1}{\sum_i p_{T,i}}
\sum_i
p_{T,i}
\begin{pmatrix}
\cos^{2}\phi_i &
\cos\phi_i\sin\phi_i\\
\cos\phi_i\sin\phi_i &
\sin^{2}\phi_i
\end{pmatrix}.
\label{eq:A2}
\end{equation}

Using the trigonometric identities

\[
\cos^{2}\phi=\frac{1+\cos2\phi}{2},
\]

and

\[
\sin^{2}\phi=\frac{1-\cos2\phi}{2},
\]

together with

\[
\sin2\phi=2\sin\phi\cos\phi,
\]

the sphericity tensor can be written as

\begin{equation}
S^{L}_{XY}
=
\frac{1}{2N\langle p_T\rangle}
\sum_i
\begin{pmatrix}
p_{T,i}(1+\cos2\phi_i) &
p_{T,i}\sin2\phi_i\\
p_{T,i}\sin2\phi_i &
p_{T,i}(1-\cos2\phi_i)
\end{pmatrix},
\label{eq:A3}
\end{equation}

where

\[
\langle p_T\rangle
=
\frac{1}{N}
\sum_i p_{T,i}.
\]

Choosing the coordinate system such that the $x$-axis coincides with the second-order symmetry plane ($\phi\rightarrow\phi-\Psi_{2}$), the reflection symmetry about the reaction plane gives

\[
\left\langle
\sin2(\phi-\Psi_2)
\right\rangle
=0.
\]

Consequently, the off-diagonal elements vanish and Eq.~(\ref{eq:A3}) reduces to

\begin{equation}
S^{L}_{XY}
=
\frac{1}{2N\langle p_T\rangle}
\sum_i
\begin{pmatrix}
p_{T,i}\left[1+\cos2(\phi_i-\Psi_2)\right] & 0\\
0 &
p_{T,i}\left[1-\cos2(\phi_i-\Psi_2)\right]
\end{pmatrix}.
\label{eq:A4}
\end{equation}

Assuming that the transverse momentum is approximately uncorrelated with the azimuthal anisotropy,

\begin{equation}
\left\langle
p_T
\cos2(\phi-\Psi_2)
\right\rangle
\simeq
\langle p_T\rangle
\left\langle
\cos2(\phi-\Psi_2)
\right\rangle,
\label{eq:A5}
\end{equation}

the sphericity tensor simplifies to

\begin{equation}
S^{L}_{XY}
\simeq
\frac12
\begin{pmatrix}
1+\langle\cos2(\phi-\Psi_2)\rangle & 0\\
0 &
1-\langle\cos2(\phi-\Psi_2)\rangle
\end{pmatrix}.
\label{eq:A6}
\end{equation}

Using the definition of the elliptic flow coefficient,

\begin{equation}
v_2
=
\left\langle
\cos2(\phi-\Psi_2)
\right\rangle,
\end{equation}

Eq.~(\ref{eq:A6}) becomes

\begin{equation}
S^{L}_{XY}
=
\frac12
\begin{pmatrix}
1+v_2 & 0\\
0 &
1-v_2
\end{pmatrix}.
\label{eq:A7}
\end{equation}

Since the matrix is diagonal, its eigenvalues are simply

\begin{equation}
\lambda_1
=
\frac12(1+v_2),
\qquad
\lambda_2
=
\frac12(1-v_2),
\label{eq:A8}
\end{equation}

with $\lambda_1>\lambda_2$ for positive elliptic flow.

The transverse sphericity is defined as

\begin{equation}
S_T
=
\frac{2\lambda_2}
{\lambda_1+\lambda_2}.
\label{eq:A9}
\end{equation}

Substituting Eq.~(\ref{eq:A8}) into Eq.~(\ref{eq:A9}) immediately gives

\begin{equation}
\boxed{
S_T
=
1-v_2.
}
\label{eq:A10}
\end{equation}

Equation~(\ref{eq:A10}) demonstrates that, under the factorization approximation, transverse sphericity decreases linearly with increasing elliptic flow. This behaviour is qualitatively consistent with the analytical relation derived for transverse spherocity, where the minimization procedure leads to a nonlinear dependence on the second-order flow harmonics.

\section{Fourier expansion of $\mathbf{|\sin x|}$}
\label{app:fourier}

The analytical relation between transverse spherocity and the flow harmonics requires the Fourier expansion of the $|\sin x|$. Since $|\sin x|$ is an even function with period $\pi$, its Fourier series contains only cosine terms with even harmonics and can be written as

\begin{equation}
|\sin x|
=
\frac{a_{0}}{2}
+
\sum_{n=1}^{\infty}
a_{n}\cos(2nx),
\label{eq:F1}
\end{equation}

where the Fourier coefficients are

\begin{equation}
a_{0}
=
\frac{1}{\pi}
\int_{-\pi}^{\pi}
|\sin x|\,dx,
\label{eq:F2}
\end{equation}

and

\begin{equation}
a_{n}
=
\frac{1}{\pi}
\int_{-\pi}^{\pi}
|\sin x|
\cos(2nx)\,dx.
\label{eq:F3}
\end{equation}

Since $|\sin x|$ is an even function,

\begin{equation}
a_{0}
=
\frac{2}{\pi}
\int_{0}^{\pi}
\sin x\,dx
=
\frac{4}{\pi},
\end{equation}

which gives

\begin{equation}
\frac{a_{0}}{2}
=
\frac{2}{\pi}.
\label{eq:F4}
\end{equation}

For the higher-order coefficients,

\begin{equation}
a_{n}
=
\frac{2}{\pi}
\int_{0}^{\pi}
\sin x
\cos(2nx)
\,dx.
\label{eq:F5}
\end{equation}

Using the trigonometric identity

\begin{equation}
\sin x\cos(2nx)
=
\frac12
\left[
\sin((2n+1)x)
-
\sin((2n-1)x)
\right],
\label{eq:F6}
\end{equation}

Eq.~(\ref{eq:F5}) becomes

\begin{equation}
a_{n}
=
\frac{1}{\pi}
\int_{0}^{\pi}
\left[
\sin((2n+1)x)
-
\sin((2n-1)x)
\right]
dx.
\label{eq:F7}
\end{equation}

Using

\begin{equation}
\int_{0}^{\pi}
\sin(kx)\,dx
=
\frac{2}{k},
\qquad
(k~\mathrm{odd}),
\label{eq:F8}
\end{equation}

one obtains

\begin{equation}
a_{n}
=
\frac{1}{\pi}
\left[
\frac{2}{2n+1}
-
\frac{2}{2n-1}
\right]
=
-\frac{4}{\pi(4n^{2}-1)}.
\label{eq:F9}
\end{equation}

Substituting Eqs.~(\ref{eq:F4}) and (\ref{eq:F9}) into Eq.~(\ref{eq:F1}), the Fourier expansion of $|\sin x|$ is

\begin{equation}
\boxed{
|\sin x|
=
\frac{2}{\pi}
-
\frac{4}{\pi}
\sum_{n=1}^{\infty}
\frac{\cos(2nx)}
{4n^{2}-1}.
}
\label{eq:F10}
\end{equation}

The appearance of only even harmonics follows directly from the $\pi$-periodicity of $|\sin x|$. This property is essential in the analytical derivation of transverse spherocity, as it implies that only the even-order flow harmonics contribute to the minimization functional.

\end{document}